\pdfoutput=1
\documentclass[9pt, table, x11names, twocolumn]{article}
\usepackage{graphicx}
\usepackage[frozencache,cachedir=.]{minted}
\usepackage{tcolorbox}
\usepackage{chngpage}
\usepackage[nottoc,notlot,notlof]{tocbibind}
\graphicspath{ {./images/} }
\usepackage[bookmarks=false]{hyperref}

\usepackage{branding}

\begin{document}

\togglecolumns 

\begin{titlepage}
    
    {\transparent{.5}\includegraphics[width=1.15\textwidth,set height=2.9cm,left=\textwidth]{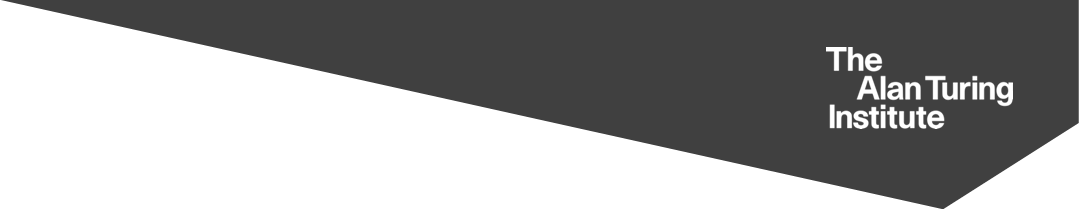}}
    
    \vspace*{1cm}
    
    \LARGE    
    \noindent Project Report: Wales Multimorbidity Machine Learning (WMML) Collaboration with The Alan Turing Institute
    
    \vspace{2cm}

    \titlecolor
    \huge
    {\fontfamily{\titlefont}\selectfont 
    
    \noindent \textbf{Developing and Publishing Code for\\
                      Trusted Research Environments:\\
                      Best Practices and Ways of Working}
        
    \vspace{1cm}
    \Large
    \noindent Ed Chalstrey, Research Data Scientist

    \noindent Research Engineering Group, The Alan Turing Institute
    
    \normalsize 
    \noindent October 2021
    }
    
    \normalsize
    \color{black}
    \vspace{3.5cm}
    \noindent Enquires:
    
    \noindent Ed Chalstrey \\
   echalstrey@turing.ac.uk
    
    \noindent The Alan Turing Institute \\
    British Library \\
    96 Euston Road \\
    London \\
    NW1 2DB
    
\end{titlepage}
\newpage

\vspace*{1cm}
\tableofcontents

\newpage



\twocolumn[ 
\begin{@twocolumnfalse} 

\begin{abstract}
  This report discusses 3 distinct, but overlapping topics. Firstly, it recommends the tools and best practices for research software engineering and data science that are most relevant to the researchers working on the Wales Multimorbidity Machine Learning (WMML) project. Secondly, it expands upon these recommendations for the specific use case of Trusted Research Environments (TREs), with development workflows for computational research in TREs offered that respect and complement existing best practices. Finally, it discusses the considerations around publishing research code that is developed to run within a TRE on sensitive data, offering practical advice that researchers using TREs can follow.
\end{abstract}
\vspace{10mm} 

\end{@twocolumnfalse}
] 

\thispagestyle{fancy} 

\section{Introduction} \label{introduction}

Trusted Research Environments (TREs) are becoming commonly used for the analysis of data from a range of sources, particularly electronic health records (EHRs). Data within TREs are kept secure and are only accessible following appropriate approvals and access being granted, to comply with the legal requirements of data providers like the National Health Service (NHS), allowing research to be carried out safely \citep{arenas_design_2019}. The volume of collaborative TRE-based research is growing, a trend that began before, but has been accelerated by, the COVID-19 pandemic.

This report summarizes a collaborative project between Swansea University, The University of Manchester and The Alan Turing Institute. The objectives of this partnership project were: (i) to develop a strategy for the Swansea and Manchester based researchers participating in the Wales Multimorbidity Machine Learning (WMML) project to follow, to implement best practices for research software engineering and data science in the context of the SAIL (Secure Anonymised Information Linkage) Databank TRE at Swansea, and (ii) to develop a strategy to move the researcher's code to a form that is citation and publication ready, outside of the TRE. This report describes the resulting recommendations from The Alan Turing Institute, which can be used as a template for making the transition from script-based (and notebook-based) research code that works on sensitive data in a TRE, to a format that is publicly accessible and can be cited by future research.

The WMML researchers are working on various novel analytical techniques and machine learning methods for clustering diseases in the Wales Multimorbidity e-Cohort (WMC) within the SAIL databank, with the aim of uncovering the multimorbidities that lead to the largest problems for both the NHS and individuals. One example of this research is \cite{rafferty_ranking_2021}. Following recommendations made in section \ref{publishing} of this report, research code associated with this paper is also published to Zenodo; see \cite{james_rafferty_2021_5285009}. The WMC is an anonymised population-scale healthcare dataset, developed within SAIL from other source data by \cite{lyons_protocol_2021}.

The recommendations discussed in this report include alternative workflows for the development of research code to be used for analysis within TREs on sensitive data, as well as best practices for working collaboratively within a TRE. In addition, recommendations are made on how to publish the work done in a TRE, including the research code. For the WMML project, this code includes the novel methods for multimorbidity clustering, as well as any scripts used for data preparation and manipulation. This report discusses how (and why) to publish this code alongside the results of the research, which include any analyses, statistics, data visualizations and other outputs relevant for publication.

This report focuses on adapting existing best practices in research software engineering and data science for the specific use case of the SAIL Databank TRE and the WMML project. However, much of the guidance developed here, in the context of a live research project, will be of broader relevance. The target audience includes any researcher or data scientist looking to work with, or publish research from a TRE. The reader is assumed to have programming skills in a language commonly used for data science such as Python or R and some familiarity with GitHub or GitLab for version controlled code projects. Resources that can help with these prerequisites are offered in section \ref{training}.

It is also hoped that this report evidences the need for computational research projects in the healthcare field to spend time and resources on implementing these recommendations. Though an operational product in the form of published research code may seem like a secondary objective, with healthcare research outputs focused more on the clinical and policy implications of results, ensuring that such code is developed to a high standard and published will increase the odds that any important results can be reproduced. This is of direct relevance to clinicians, policymakers and other researchers who are making decisions based on those results and need to understand how reliable the findings are.

\newpage
\subsection{Project Scoping}

The project that has culminated in this report was conducted by (and report authored by) Ed Chalstrey on behalf of The Alan Turing Institute, in collaboration with WMML researchers and principal investigators (PIs), see \ref{acknowledgements} (Acknowledgements). Scoping for the report involved individual and group discussions between the author and WMML team members, carried out over the duration of the project via conference call and email. Initial discussions revolved around defining the project output, suggestions for which included code contributions to \citep{james_rafferty_2021_5285009} and other WMML code repositories within SAIL.

Upon further discussion, it became apparent that much of the advice presented in \href{https://the-turing-way.netlify.app/welcome.html}{The Turing Way} \citep{the_turing_way_community_2019_3233986}, an open source handbook for reproducible, ethical and collaborative data science, developed in large part by members of The Alan Turing Institute, would be of direct relevance to WMML researchers. However, recommendations of this nature that cater to TRE-based research specifically are not something currently included in the handbook or comparable resources elsewhere. This gap was identified as something valuable to address, both for the WMML project, for future SAIL projects and for TRE-based research in general.

The author and collaborators coalesced around the idea of addressing  the gap in the literature with this report, which combines ways of working suggestions of direct relevance to the WMML project with more generally applicable guidance for the publishing of research conducted with TREs. This output was suggested by the author and independently agreed upon each of the collaborators. Useful revisions to this report have been suggested by the collaborators listed in \ref{acknowledgements} (Acknowledgements) and implemented by the author.

\begin{tcolorbox}[colback=white!95!blue, colframe=DodgerBlue3]
\textbf{Turing-WMML Collaboration Objectives:}
\begin{itemize}
    \item Offer best practices for research software development in the TRE context
    \item Provide a template for publishing code from research carried out with TREs
    \item Ensure WMML project team can realise the first two objectives
    \item Report summarising collaboration and resulting recommendations for TRE research
\end{itemize}
\end{tcolorbox}

\section{The Turing Way for TREs} \label{turingwaytres}

This section of the report, discusses some of the most pertinent best practices for reproducible data science and scientific software development in the context of TRE-based research. Much of this section is directly adapted from \href{https://the-turing-way.netlify.app/welcome.html}{The Turing Way} handbook \citep{the_turing_way_community_2019_3233986}. A variety of topics applicable to TRE-based research are covered including: \ref{notebooks}) using notebooks (such as Jupyter or RMarkdown), \ref{testing}) how to ensure that code meets the quality requirements necessary for publishing research code (via testing, formatting and use of continuous integration) and \ref{environments}) controlling the computational environment to aid reproducibility. Advice on how to best utilise each of these are offered alongside commentary on the pitfalls and tradeoffs of various approaches.

This report assumes a basic degree of familiarity with the core concepts of version control with Git. For an overview of this topic in the context of reproducible scientific research code, see the \href{https://the-turing-way.netlify.app/reproducible-research/vcs.html}{"Version Control" subsection} of the "Guide for Reproducible Research" within The Turing Way online handbook \citep{the_turing_way_community_2019_3233986}. See also section \ref{training} of this report for additional resources and relevant training.

\subsection{Using Notebooks Effectively} \label{notebooks}

It's common for researchers to use digital notebooks such as Jupyter or RMarkdown for carrying out data science tasks, especially when this involves visualisation, and this is no different when these tools are available within TREs. When publishing research nowadays, it's not unusual to include figures than began life as the output of a data analysis task in a notebook. Indeed notebooks themselves have in some cases become the primary output of computational research in the form of executable papers, published in open repositories and hosted with online solutions such as Binder \citep{lasser_creating_2020}. The applicability of such tools to the publishing of TRE-based research outputs will be discussed later in this report (see section \ref{executable}), but here several suggestions are made about how best to utilise notebooks themselves in both the context of TRE research and the desire to publish code as a research output.

It's important to recognise that whilst notebooks are invaluable resources for experimentation and reporting, they do not fulfil the same purpose as afforded by scripts in version-controlled repositories. Notebooks make conflict resolution for version control difficult and each time they are opened, cells can be run in different orders, meaning that it can be easy to lose track of what code has already been run. Whilst it is possible to keep notebooks in version-controlled repositories (hosted on GitLab for example), the code cells of Jupyter notebooks in particular are not ideal places to store important methods code for machine learning projects such as the WMML. The main reason for this is that the Jupyter notebook files (.ipynb extension) are JSON documents that include metadata. RMarkdown notebooks by contrast are based on markdown documents, which are better for tracking changes. Nonetheless, both options are suboptimal for the purpose of developing publishable research code of a high quality (unless the aim is to publish the notebook itself as an executable paper, which will be discussed later on in this report). This is because they lack the functionality to test and format methods code (a topic covered in section \ref{testing}).

In addition, inclusion of methods code will quickly render notebooks unreadable, defeating a key purpose for the usage of notebooks in the first place. Core research methods code should be developed separately and imported into notebooks if and when needed for analysis.

In the context of TRE-based research, notebooks can be used as a tool to explore and report on sensitive data, rather than as a research output themselves for export from the TRE (though exceptions to this will be discussed in section \ref{tredevworkflows}). Notebooks could for instance contain sensitive data in the commit history of their GitHub/GitLab repository, even if the data is not present in the code cells at the time of export.

Having said this, the combination of code cells that can generate plots for figures and markdown cells for text can make notebooks a useful starting point for the structuring of a research paper. Using tools readily available within the SAIL TRE, it is relatively simple to convert an interactive notebook to a static research paper PDF. Indeed, several members of the WMML team have found this approach fruitful already. In order to do this, pandoc (available at \href{https://pandoc.org/}{pandoc.org}) must be installed (as is the case in SAIL).

To convert a Jupyter notebook to PDF, simply use nbconvert as follows:
\begin{minted}{python}
jupyter nbconvert --to pdf notebook.ipynb
\end{minted}

Converting an R markdown notebook to PDF, is even easier when using the RStudio application; simply choose the "Knit to PDF" option.

In both cases, there are options to embed LaTeX and include useful features such as a table of contents when doing this conversion (documentation for this is readily available online). If notebooks are the method that researchers working in a TRE wish to use to draft their papers, converting to a static format such as PDF makes the checks required for exporting from the TRE far simpler than would be the case if exporting notebooks themselves. Documentation for nbconvert can be found at \href{https://nbconvert.readthedocs.io/}{nbconvert.readthedocs.io}, including how use cell tags to hide input (or both input and output) in generated documents, something that could be useful if generating a research paper from a notebook.

However, members of the WMML team have found that converting to PDF with nbconvert can result in errors for complex Jupyter notebooks that are hard to debug. Therefore it would be prudent to test conversion works as expected at regular intervals if generating a research paper from a notebook. A suggested approach when such errors do occur, is to generate a LaTeX source file with nbconvert, which enables easier debugging.

It's also possible to use nbconvert to convert a Jupyter notebook into a code script, which includes the content of markdown cells as comments. This can be useful if development that has been initiated in a notebook reaches a point where it would be better to continue in a script, for example if any functions have been created in code cells that need to be tested (see section \ref{testing} for more information on code testing).

Conversion of a Jupyter notebook to a script can be done like so:
\begin{minted}{python}
jupyter nbconvert --to script notebook.ipynb
\end{minted}

For the development of research methods code used in TREs, it makes sense to use scripts or modules that are independent of any notebook, especially if such code is destined for export from the TRE. Any code being developed in the context of a notebook cell that cannot be quickly re-written should be moved to a separate file that is tracked within the Git repository (e.g. a Python module or R script). Developing research methods within a Git repository offers more advantages than just version control. It also increases the ease with which the code can be developed to a standard of quality appropriate for both publishability and reproducibility, via the use of tests and formatting amongst other things (see \ref{testing}).

\begin{tcolorbox}[colback=white!95!blue, colframe=DodgerBlue3]
\textbf{(\ref{notebooks}) Key Points:}
\begin{itemize}
    \item Core research code should be developed in version controlled repositories
    \item Jupyter/Rmd notebooks should not be exported from the TRE
    \item Notebooks containing draft papers and figures can be converted to static PDFs
\end{itemize}
\end{tcolorbox}

\subsection{Code Testing and Quality} \label{testing}

Test-driven development is a software engineering philosophy that prioritises the writing of unit tests, which test individual elements of code rather than entire applications, before the writing of the application code itself \citep{the_turing_way_community_2019_3233986}. In the context of research software, unit testing (and by extension, test-driven development) has several benefits. Testing makes development faster because bugs are far easier to isolate and can be spotted early. By encouraging the modularisation of the research code into functions and groups of functions, researchers are better able to focus on the lines of code relevant to the part of their analysis being worked on \citep{the_turing_way_community_2019_3233986}. Researchers who set up tests for their functions can be confident those functions work as expected when the tests pass. This means functions can be easily reused whenever needed. When working on analyses in notebooks, these functions can be imported from the local script or module where development is taking place (or from the package if the code has already been packaged).

The recommended testing framework for Python is "pytest" and for R, there are several good choices such as "testthat", "tinytest" and "svUnit". Each of these are available on PyPI and CRAN respectively (meaning they are installable in SAIL via pip or install.packages).

In the context that this report seeks to address (that of publishing TRE-based research code), having well tested (and documented) code will ensure that researchers can prioritise the minimal amount of code for export from the environment (assuming development takes place within the TRE, see section \ref{tredevworkflows} for more on development options). This reduces the burden on the export process and should save time. Furthermore, writing research code that is modularised, well documented and appropriately tested, ensures that results generated with that code can be reproduced. It also adds to the clarity of the code when published, better enabling paper reviewers and other scientists that access, run or wish to further develop the code, to understand what it does and how it works.

Testing of software can be automated with continuous integration (CI). In software development, CI is typically used when multiple people are working on the same codebase and want to push changes regularly without breaking anything. It can however also be a valuable tool for individual researchers looking to produce well structured and well tested code for publication. In addition, by forcing the researcher to specify the requirements for their code to run (e.g. the packages/libraries and their versions), CI can help to keep track of the computational environment. The computational enviroment is discussed further in section \ref{environments}.

CI can be added to both GitLab and GitHub repositories and be set up so that unit tests (and other checks) are run each time a feature or bug-fix branch is merged into the main code. The advantage of this approach is that breaking changes in the code should be found early, because they cause (well-written) tests to fail. CI can be set up such that merging into the main code branch is not possible without tests and other checks passing on the new code branch. This makes bugs easier to spot and prevents stable code branches becoming compromised \citep{the_turing_way_community_2019_3233986}.

To make effective use of CI, a good fundamental understanding of version control is required. Researchers that don't have this experience, can read the subsection titled \href{https://the-turing-way.netlify.app/reproducible-research/vcs.html}{"Version Control"} in the "Guide for Reproducible Research" section of The Turing Way online handbook. There is also a subsection of the handbook on \href{https://the-turing-way.netlify.app/reproducible-research/ci.html}{CI}, which details how to get set up with GitHub actions. Documentation on how to use CI in GitLab is available at \href{https://docs.gitlab.com/ee/ci/}{docs.gitlab.com/ee/ci/}, however it is worth noting that at the time of writing, CI tools are not included for GitLab in SAIL. In order to make use of CI as part of the internal TRE development workflow described later on in section \ref{internal} of this report, GitLab CI will need available in the TRE.

One other way of improving the quality of research code destined for publication is to adopt a consistent code style. A code style is a set of conventions on how to format code, for example standardising things like indentation and the placement of comments. Conforming to a particular code style can make code easier to understand for collaborators and reviewers. Code style checks can also be set up to run alongside tests with CI in an automated fashion, enabling consistency throughout the development process. Once again, \href{https://the-turing-way.netlify.app/reproducible-research/code-quality/code-quality-style.html}{The Turing Way} contains additional information on this topic, including recommended formatter tools for Python, R and other languages. 

\begin{tcolorbox}[colback=white!95!blue, colframe=DodgerBlue3]
\textbf{(\ref{testing}) Key Points:}
\begin{itemize}
    \item Testing research code speeds up development because bugs are easier to spot
    \item Automatically run tests when changes are made via Continuous Integration (CI)
    \item Consistent code styling can keep code easy to understand
\end{itemize}
\end{tcolorbox}

\newpage
\subsection{Reproducible Environments} \label{environments}

Every computer or virtual machine has a unique computational environment consisting of its operating system and the installed software (including language and package versions) and TREs are no different. When publishing research code, or indeed any scientific output, being able to reproduce results is of paramount importance to enable other researchers to check the claims made by the research. An additional consideration for computer science research is the computational environment in which published research results were generated. Capturing important aspects of this environment can ensure that others can run the code and reproduce the results \citep{the_turing_way_community_2019_3233986}.

When considering to what extent a computational environment needs to be captured, a key decision to make is whether managing the software and versions via a package management system is sufficient, or whether it is worth controlling the entire operating system, via containerisation. Containers are essentially lightweight virtual machines, which can contain their own files, software and settings and are particularly useful if projects need to run on high-performance computing (HPC) environments \citep{the_turing_way_community_2019_3233986}.

There are a variety of package management solutions for both Python and R, but this report focuses on "Conda". Conda is an open source package and environment management system that installs, runs and updates packages and their dependencies. Though Conda was originally created for Python programs, it can be used with most commonly used data science languages, including R (Anaconda, 2016). In SAIL, Conda is installed by default.

Conda can be used to create a virtual environment for a project, which keeps track of the packages (and package versions) being used by the code under development. This is possible via the "environment.yml" file, which can be used to quickly re-create the virtual environment from scratch. By including this file in their GitHub/GitLab repository, researchers can ensure that when others wish to use their project code, installing all the packages and dependencies required for the code to work is simple.

A Conda virtual environment can be created from the "environment.yml" and activated by running the following shell commands:

\begin{minted}[gobble=2]{bash}
    >   conda env create -f environment.yml
    >   conda activate projectvenv
\end{minted}

\newpage

For a Python project, the "environment.yml" file should look something like this example (which will create a virtual environment called "projectvenv"):

\begin{tcolorbox}[colback=white!95!black, colframe=white]
\begin{tcolorbox}[colback=white!95!black, colframe=white!95!black]   
\begin{minted}[
    gobble=2,
    linenos
  ]{yaml}
    name: projectvenv
    channels:
      - defaults
    dependencies:
      - numpy>=1.18
      - scipy
      - pandas
      - pip=21.2.4
      - pytest
      - python=3.9.7
\end{minted}
\end{tcolorbox}
\end{tcolorbox}

If a researcher is already using a conda virtual environment for package management, but hasn't yet created an "environment.yml", it can be created from the existing virtual environment:

\begin{minted}[gobble=2]{bash}
    >   conda env export > environment.yml
\end{minted}

New packages can also be added to the environment with a single command. For example, a researcher working in R could use the following command to add the "ggplot2" package to the environment:

\begin{minted}[gobble=2]{bash}
    >   conda install r-ggplot2
\end{minted}

Adding packages this way could be useful when experimenting with new software dependencies for research code, in advance of deciding they are essential and should be added to the "environment.yml".

For more information on how to use conda for package management, consult the \href{https://docs.conda.io/projects/conda/en/master/user-guide/tasks/manage-environments.html#}{online documentation} (see also \href{https://docs.anaconda.com/anaconda/user-guide/tasks/using-r-language/}{R specific information}). There are of course other package management solutions for both languages, and researchers should feel free to use those that duplicate (or exceed) the functionality of Conda. For R in particular, package/environment managers such as "packrat" and "renv" may be more commonly used. However this report recommends Conda, on the basis of its simplicity to get started with, its ability to build environments from a single file and to re-build them on other machines with that file.

When it comes to the question of whether to make use of containerisation, this likely falls outside the requirements of the WMML project. Containers are best placed for distributing software that needs to be run on a variety of different computers, where those computers only have to have the container software itself (e.g. Docker) installed as a pre-requisite. Containers not only include the operating system, but often also the project files (including data), neither of which seem necessary or appropriate for the WMML project's goal of publishing machine learning methods code for R and Python. For interested readers, the most popular containerisation software is \href{www.docker.com}{Docker}, however there is also \href{https://sylabs.io/guides/3.5/user-guide/introduction.html}{Singularity} which specialises in containers for HPC.

\begin{tcolorbox}[colback=white!95!blue, colframe=DodgerBlue3]
\textbf{(\ref{environments}) Key Points:}
\begin{itemize}
    \item Controlling the computational environment of a project aids reproducibility
    \item Packages and versions for Python/R code can be managed with Conda
    \item Containers can reproduce the entire computational environment if needed
\end{itemize}
\end{tcolorbox}

\subsection{Reproducible Workflows} \label{reproducibleworkflows}

In scientific research, and computer science in particular, we often want to ensure that the results of our analyses are \emph{reproducible}. Reproducibility in scientific research can be differentiated from \emph{replicability}, whereby consistent results and similar conclusions are reached across different studies and methodologies answering the same research question.

A fully reproducible workflow for computational research would include the option to use the same input data, computational environment and code (research methods) in order to reproduce the results exactly. 

Achieving a fully reproducible workflow is difficult in practice and working in a TRE on a dataset that isn't publicly available adds additional challenges. Most obviously, there is the question of how a paper reviewer (or other researcher) might gain access to the input data, given that the researchers themselves required a login to the TRE system for access. There is already a process in place to enable journal reviewers access to the SAIL databank, should they require it and SAIL is open to all researchers, subject to approval and adequate qualification.

However, when it comes to reproducibility from the perspective of the broader scientific community, access to data available within a TRE (such as the WMC dataset used by the WMML project in SAIL) is not immediately available. In section \ref{tredevworkflows} of this report, possible workflows for developing and publishing code from research with TREs are discussed and section \ref{synthetic} specifically addresses the possibility of sharing synthetic data as a substitute for sensitive datasets that cannot be exported from TREs.

\begin{tcolorbox}[colback=white!95!blue, colframe=DodgerBlue3]
\textbf{(\ref{reproducibleworkflows}) Key Point:} Changes in any of the following can lead to differences in the results of scientific research:
\begin{itemize}
    \item Data
    \item Code
    \item The computational environment
\end{itemize}
\end{tcolorbox}

\section{TRE Development Workflows} \label{tredevworkflows}

When it comes to the question of how to publish research conducted with sensitive data in a TRE, and in particular how to publish code, there is little in the way of best practice suggestions or guidance that has been previously written up in a comprehensive manner. Importantly, thinking about this problem forces us to tackle the question of how best to develop code for TRE research in the first place, given the appropriate constraints when it comes to accessing data within TREs.

The development of research code for use on sensitive data within a TRE could involve the development of novel methods and models, or it could merely involve the application of existing models, methods or libraries to the research question at hand. In the case of the WMML project, researchers are working in novel methods, which necessitates some of the considerations already discussed in this report, such as testing. Even were this not the case, computational research will generally involve some code scripts that can be published alongside the results of the research, to clearly demonstrate the methodology and improve reproducibility.

When deciding which parts of the research code are relevant for publication, a decision should be made to omit any code that is specific to the particular TRE. However, it is recommended that where possible, such code is at least made available to others using the same TRE. For example, a script written by a WMML researcher that could have value to other SAIL databank users, but isn't useful outside the TRE, could be stored in a shared folder, to which users are granted access by default. This will ensure that such code is available to researchers on other projects, even if not relevant to the wider research community who lack access to the TRE.

This report presents two alternative workflows for the development and publication of code from TRE-based research. The first, described in section \ref{external} and shown in figure \ref{fig:extdev}, takes the approach of developing the primary research code (that which is destined for publication) externally from a TRE, whereas section \ref{internal} and figure \ref{fig:intdev} document a workflow based on developing code within a TRE. In each workflow, researchers export their results (which could include analyses, statistics, data visualizations and reports) from the TRE for publication, but where they differ is whether the code used to generate those results also needs to be exported.

The benefits and tradeoffs of these development workflows are discussed and some light suggestions are made as to which approaches are most appropriate for the WMML project and SAIL. Regardless of the chosen development workflow, researchers on the project are assumed to have access to the TRE (including the contained sensitive data) as well as any code or results they produce.

Also highlighted are several optional extras that can further enhance the reproducibility and find-ability of the research, such as making use of synthetic data (\ref{synthetic}) and executable papers (\ref{executable}).

\subsection{External Development} \label{external}

One problem that arises from the desire to publish research code developed in a TRE, is that the burden of checking the code for inappropriate content by disclosure control reviewers (during releases from a TRE), increases with the scale and complexity of the code base. Such content could include data-disclosive comments in the code or extracts of data embedded within the code.

In addition, researchers are limited by the software tools available in the TRE for development. As an example, a limitation highlighted by discussions with the WMML team was the lack of CI tools available for GitLab within SAIL. As such, some members of the team have taken the approach to develop much of their primary research code in an external GitHub repository, in order to make use of GitHub Actions CI. Developing in this way, researchers can import snapshots of their code into the environment whenever changes have been made and an analysis needs to be re-run.

Another issue is that as virtual machines, TRE systems have the potential for slow responsiveness that is anathema to software development.

Based upon discussions between colleagues at The Alan Turing Institute, it appears that problems of the nature faced by the WMML team are common in TRE-based research. Given these issues, this report further develops the idea of working on primary research code in an external repository as a workflow for TRE research (figure \ref{fig:extdev}) and offers a comparable alternative workflow that involves developing within a TRE (figure \ref{fig:intdev}).

The external development workflow begins with the researcher creating a publicly hosted code repository (e.g. with GitHub). This repository houses the primary research methods code and can optionally be set up to include tests and CI (see section \ref{environments} of this report). Development of the code is carried out here and each time a new feature is added that will affect the researcher's analysis, the researcher can import a snapshot of the code (e.g. the most recent commit) to the TRE and re-run that analysis. This process iterates until the researcher has completed their work in the TRE and has a final analysis that is ready for publication. In this workflow, export from the TRE will only include the results of the analysis (e.g. plots generated or a draft paper); the code repository is already publicly available (section \ref{citation} will outline additional steps to make the repository citation/publication ready).

A disadvantage of external development could be that writing the code (e.g. machine learning methods for the WMML project) is difficult without access to the data it is designed to run on. It could be possible to overcome this obstacle by including dummy data, which replicates the structure of the original data from the TRE (but contains randomly generated data), in the external repository. Section \ref{synthetic}, describes in more detail how algorithmicly generated synthetic data (that is statistically representative of the original data) could also be used.

\begin{tcolorbox}[colback=white!95!blue, colframe=DodgerBlue3]
\textbf{(\ref{external}) External Development Advantages:}
\begin{itemize}
    \item Make use of development tools unavailable in the TRE such as GitHub Actions
    \item Fewer steps required to publish code developed on a public website than from a TRE
\end{itemize}

\textbf{(\ref{external}) External Development Disadvantages:}
\begin{itemize}
    \item Progress may be slowed if data access is critical for development
\end{itemize}
\end{tcolorbox}

\subsection{Internal Development} \label{internal}

Downsides of developing research code inside the TRE that have been discussed include being limited by the TRE's available software tools and the potential for a high burden on disclosure control reviewers. However, some researchers may find the convenience of developing their research code in the same environment they are running it, outweighs these downsides. For this case, figure \ref{fig:intdev} shows a modified version of the workflow where the primary research code development occurs within the TRE.

In the internal development workflow, researchers set up their repository in much the same way as the external workflow, but make use of the TREs version control system (GitLab in the case of SAIL) instead of an external repository. When the researcher has completed their work, export from the TRE will include both the research results and the code used to generate them.

An important recommendation here is to only export the final version of the code, rather than the entire repository, as this will simplify the audit process for export. Checking an entire commit history for sensitive data leaks will be far more time consuming than doing so for just the code itself.

\begin{figure}[]
    \hspace*{-1.7cm}
    \includegraphics[width=0.6\textwidth]{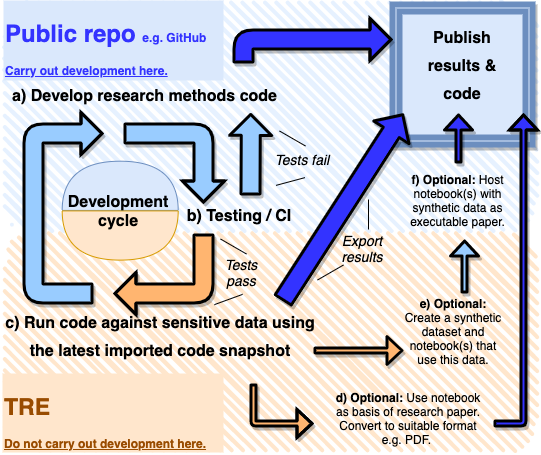}
    \caption{Workflow for external development of TRE research code. Snapshots of externally developed code are imported into the TRE whenever new features are added that can be used for the analysis of sensitive data. Only the results of the analysis need be exported from the TRE, since externally developed code is ready for publication. The blue backdrop and arrows show steps involving and moving into a public code repository, whereas orange represents the TRE. Steps indicated by darker arrows occur later in the workflow. Smaller arrows link to optional steps that can be taken to further improve the reproducibility and availability of the research.}
    
    \label{fig:extdev}
    
\end{figure}

\begin{figure}[]
    \includegraphics[width=0.6\textwidth]{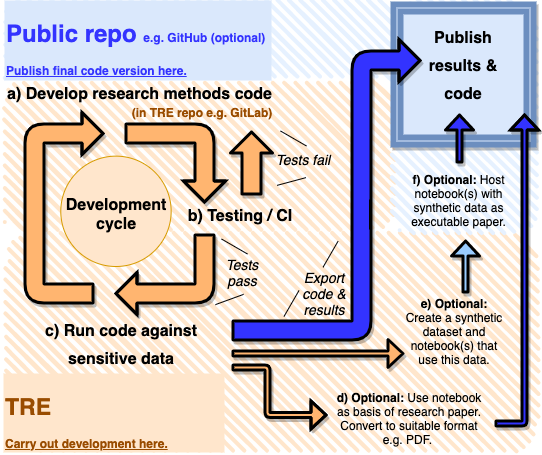}
    
    \caption{Workflow for internal development of TRE research code. A snapshot of the research code (e.g. the final GitLab commit) is exported alongside the results of the analysis, after passing any security processes required to export from the TRE. The blue backdrop and arrows show steps involving and moving into a public code repository (or other online location to which the code will be uploaded for publishing purposes), whereas orange represents the TRE. Steps indicated by darker arrows occur later in the workflow. Smaller arrows link to optional steps that can be taken to further improve the reproducibility and availability of the research.}
    
    \label{fig:intdev}
    
\end{figure}

Figures \ref{fig:extdev} and \ref{fig:intdev} also highlight several optional steps aimed at improving the reproducibility of the research including making use of synthetic data and executable research papers, which are discussed in sections \ref{synthetic} and \ref{executable} respectively.

\begin{tcolorbox}[colback=white!95!blue, colframe=DodgerBlue3]
\textbf{(\ref{internal}) Internal Development Advantages:}
\begin{itemize}
    \item Code can be developed in the same environment it is run
\end{itemize}

\textbf{(\ref{internal}) Internal Development Disadvantages:}
\begin{itemize}
    \item Useful development tools maybe unavailable in the TRE e.g. GitLab CI in SAIL
    \item Exporting code from a TRE can run the risk of accidental data disclosure
\end{itemize}
\end{tcolorbox}


\subsection{Synthetic Data} \label{synthetic}

As previously stated, one of the problems with reproducibility of research conducted with data stored securely in a TRE, is that since that data is not publicly available, other researchers wishing to reproduce the results of the research are unable to do so without first gaining access. There is however a partial substitute for this in the form of creating publicly available synthetic data resembling the original data, which the published research code can also be run on.

Synthetic data can be described as data generated by a statistical model which replaces identifying or sensitive values from an original confidential dataset with draws from the model \citep{drechsler_empirical_2011}. Synthetic data should retain some of the statistical properties of the original data, without leaking the sensitive original data itself. There are a variety of methods and software packages that have been developed for this task. Indeed, synthetic data research is an active field in of itself. A key challenge these methods face is balancing the utility of the synthetic data as a substitute for the original dataset, against the risk of accidental disclosure of sensitive data \citep{drechsler_empirical_2011}.

In the context of TRE-based research, it will often be the case that the very reason for working in a TRE is to analyse datasets that are inherently too sensitive to publish publicly, as is the case with the WMC in SAIL being analysed by the WMML project team. In cases such as this, the question then is: can a synthetic version of the dataset be produced that can be published alongside the code and analysis results? Doing this would enable reviewers and other researchers to try out the published methods (the code) for themselves on the synthetic dataset, with the caveat that they will not be able to reproduce the exact results of the code being run on the original data.

Figures \ref{fig:extdev} (e) and \ref{fig:intdev} (e) show how synthetic data could be incorporated into a TRE development workflow. In addition to its application in improving research reproducibility from the perspective of external researchers, synthetic data could also be utilised by researchers who are waiting for access to a TRE when beginning a project. This would enable them to make a start on their research before gaining access.

For the purposes of the WMML project, it's unlikely that creating a synthetic version of the WMC would prove a worthwhile endeavour, given the maturity of the project and the challenges associated with creating non-disclosive synthetic data. This may however be a route work looking into for future SAIL TRE research projects that aim to publish research carried out in the TRE. A good starting point would be to investigate how existing software packages for synthetic data generation (such as \href{https://www.synthpop.org.uk/}{synthpop}) perform on the healthcare data being analysed in SAIL. Researchers should also investigate any existing synthetic data projects already in progress in the SAIL databank.

A more lightweight approach that could be of value to the WMML project and others where synthetic data research falls out of scope, is to produce a small dummy dataset that has the same data structure (e.g. table columns and data types) as the original sensitive dataset, but randomly generated data. A dummy dataset of this nature could be small enough to be included as part of the published code repository and would give an example of the kind of data that the methods code could be run on, thereby reducing at least some of the friction to reproducibility. Where working in notebooks, a duplicate version of the notebook that loads the dummy data instead of the original dataset could be included (see figures \ref{fig:extdev} e and \ref{fig:intdev} e).

\begin{tcolorbox}[colback=white!95!blue, colframe=DodgerBlue3]
\textbf{(\ref{synthetic}) Key Points:}
\begin{itemize}
    \item Project researchers could develop code on synthetic data prior to TRE access
    \item Dummy data replicating the original data structure can be published with code
    \item Synthetic data that retain original data size and properties are a larger challenge
    \item External researchers lacking TRE access could further develop published code that has accompanying dummy/synthetic data
\end{itemize}
\end{tcolorbox}

\subsection{Executable Research Papers} \label{executable}

In section \ref{notebooks}, this report cautioned against the exporting of notebooks from TREs. However, one exception could be if a duplicate of a notebook used for the researchers analysis is created that loads a synthetic or dummy dataset, suitable for export (see figures \ref{fig:extdev} and \ref{fig:intdev} e). If a researcher has reached the point where they're publishing notebooks and code to online repositories, they're just one step away from publishing an interactive online executable research notebook. These online notebooks demonstrate research methods as clearly as is possible, by actually running them, something that is fast becoming the gold standard of reproducible computational research. By extension, notebooks that contain an entire research paper can be hosted as executable papers \citep{lasser_creating_2020}.

Pairing executable papers with a published synthetic dataset could be a good option for achieving the highest degree of scientific reproducibility possible for sensitive data research (including TRE-based research). Figures \ref{fig:extdev} (f) and \ref{fig:intdev} (f) show how executable papers could fit in to the TRE research development workflow.

A key software tool used for creating executable papers from Jupyter notebooks is Binder. Binder uses a tool called "repo2docker" to create a Docker image (the instructions to build a Docker container) for a GitHub repository, based on included configuration files (such as Conda's "environment.yml" file). The resulting image can be accessed via a cloud-based BinderHub, which allows anyone with the url to view, edit and run the Jupyter notebook from their web browser \citep{the_turing_way_community_2019_3233986}. Check out \href{https://mybinder.readthedocs.io/en/latest/examples/sample_repos.html#conda-environment-with-environment-yml}{this example} of a Python Jupyter notebook hosted with Binder which uses packages specified by a Conda "environment.yml" file.

Hosting a research notebook online in this way gives people the ability to see the results of analyses in the notebook reproduced in real time, as well as the ability to edit the notebook code to try out different parameters or variations of the analysis. Binder can also be used with R, but at the time of writing, lacks the capability to create a url that loads an RMarkdown notebook in the same way as for Jupyter (note: Binder is an open source community project that is constantly evolving and readers should investigate current options for R notebooks online). To learn more about Binder, review the relevant section in \href{https://the-turing-way.netlify.app/reproducible-research/renv/renv-binder.html}{The Turing Way handbook}.

Executable papers can be used to demonstrate that choices made when carrying out the research only had a minor impact on the outcome. This approach both strengthens the conclusions from the readers perspective and allows author to feel more confident about the decisions made during their analysis \citep{lasser_creating_2020}.

It's likely that producing online executable versions of the research notebooks being developed by WMML researchers falls outside the achievable goals within the time frame of the project. However, researchers working on TRE-based projects should not feel discouraged from pursuing this option in future, in tandem with the publishing of a synthetic dataset (or dummy dataset - see section \ref{synthetic}) that can be loaded by the executable paper.

A potential blocker to this approach of interactive research communication, is that some analyses carried out in notebooks may require large or specific compute resources that aren't feasible in the context of a Binder notebook. That being said, if such a notebook was set up to run on a small synthetic dataset, this could reduce the barrier presented by lesser compute resources.

\begin{tcolorbox}[colback=white!95!blue, colframe=DodgerBlue3]
\textbf{(\ref{executable}) Key Points:}
\begin{itemize}
    \item Binder can be used to create online notebooks anyone with the url can access
    \item For sensitive data research, executable papers could be paired with synthetic data
\end{itemize}
\end{tcolorbox}

\section{Publishing Research Outputs} \label{publishing}

At the outset of this collaboration project with The Alan Turing Institute, the WMML project team expressed the desire to investigate best practices for conducting and publishing reproducible research, in particular the research code, in the specific context of working with a TRE. In this report, workflows for TRE research and development are described which include a final step of publishing the research (see the final stage of figures 1 and 2). This final step in both example workflows involves making the code publicly available alongside the paper, to enhance the reproducibility of the research.

This section includes recommendations on how the publicly available research code can be made easily citable and findable, as well as how to choose an appropriate software license.

\subsection{Citation Ready Software} \label{citation}

There are various reasons why people don't cite the research software they use, but one of the biggest ones is that it’s not clear how. Following the suggestions outlined here will maximise the chance of the WMML researchers published code being cited correctly, allowing others to build on the work in a way that acknowledges their contribution. In addition, it will increase the odds of researchers working on machine learning methods or multimorbidity from discovering their work.

A key suggestion here is to create a Digital Object Identifier (DOI) for the code. DOIs are unique identifiers or persistent links for digital objects (including data as well as code) that make it much easier for researchers to cite each other's work. DOIs reduce the risk of link rot and mean researchers can track how their work is being used and cited \citep{the_turing_way_community_2019_3233986}. A DOI for the research code can be generated by uploading it to \href{https://zenodo.org/}{Zenodo}.

In addition, researchers can sign up for an \href{https://orcid.org/login}{ORCID} account. An ORCID (Open Researcher and Contributor ID) allows a researcher to provide a unique identity for their body of work independent of their name. It enables them to collect together (and others to find) all their research papers and related outputs so they can be easily cited \citep{the_turing_way_community_2019_3233986}.

Another useful option is to inlcude a \href{https://citation-file-format.github.io/}{CITATION.cff file} in the published code's GitHub repository, containing a message explaining how someone can refer to different outputs from the research. This file can include the DOI, as well as a link to the researcher's ORCID account and other relevant metadata.

\begin{tcolorbox}[colback=white!95!blue, colframe=DodgerBlue3]
\textbf{(\ref{citation}) Key Suggestions:}
\begin{itemize}
    \item Upload code to Zenodo to get a unique Digital Object Identifier (DOI)
    \item Sign up for an Open Researcher and Contributor ID (ORCID)
    \item Include a citation file alongside the code when published in a GitHub repo
\end{itemize}
\end{tcolorbox}

\subsection{Packaging Software} \label{packaging}

As an optional extra step, researchers may wish to convert their software into a package. This
allows for the option of publishing the code to online repositories such as PyPI for Python and CRAN for R, increasing the ease with which this code can be re-used in future projects. This could be of value in TREs such as SAIL, which has open ports to PyPI and CRAN servers, enabling easy install of research code developed in past projects to new work-spaces or environments.

Explaining the process of turning a code repository into, for example, a "pip" installable Python package on PyPI falls outside the scope of this report. However, research projects following the TRE development workflows described in section \ref{tredevworkflows} of this report should be able to create packages from their software, by modifying these example workflows to include the setting up of their code repositiories with any required files and publishing the code to the chosen package repository as a final step.

\begin{tcolorbox}[colback=white!95!blue, colframe=DodgerBlue3]
\textbf{(\ref{packaging}) Key Point:} Packaging research code increases ease of re-use in future projects
\end{tcolorbox}

\subsection{Software Licenses} \label{licenses}

When publishing software, its common practice to include a license which governs the extent of use or redistribution of the software. The easiest way to choose the license most appropriate for a piece of research software is to visit \href{https://choosealicense.com/}{choosealicense.com}, which offers a straightforward mechanism to help pick one. To learn more about software licenses before making a choice, consult the licensing section in \href{https://the-turing-way.netlify.app/reproducible-research/licensing.html}{The Turing Way} which covers this in more detail \citep{the_turing_way_community_2019_3233986}.

For maximal engagement with published code, it is recommended to use a permissive open source software licence. However, some projects may find it necessary to opt for a less permissive license if their research code requires propriety software tools. It is recommended to check out the software licence suggestions of the funder(s) of research, but in general, charitable or altruistic funders are more likely to encourage open source licenses.

Once a licence has been chosen, it can be saved to a file within a public code repository (e.g. GitHub) alongside the research code. This file can be named appropriately (e.g. License.txt or License.md) and stored it in the top level of the repository directory structure so it's easy for people to find.

\begin{tcolorbox}[colback=white!95!blue, colframe=DodgerBlue3]
\textbf{(\ref{licenses}) Key Suggestion:} Choose an appropriate software licence for published research code and include it in any public repositories that contain the code
\end{tcolorbox}

\section{Conclusion} \label{summary}

The volume of TRE-based research is likely to grow in the coming years and decades, so the establishment of a set of best practices and norms for developing and publishing research code for these projects should be of broad appeal to the research community. This applies in particular at the intersection of healthcare and data science. As such, this report should be viewed as an initial step towards formalising these processes and not the final word on the topic.

By developing code to a common workflow such as described in this report, the WMML project (as well as other SAIL projects and projects in other TREs) will be able to increase the efficiency of collaboration within a project, maximise the visibility of the research once published and the improve the reproducibility of the results.

\begin{tcolorbox}[colback=white!95!orange, colframe=orange]
\textbf{Report Summary:}
\begin{itemize}
    \item Develop methods code as scripts/modules in a GitLab/GitHub repository
    \item Developing code in a public repository external from the TRE should be considered
    \item Test code to ensure it works as expected
    \item Where possible, automate tests and code quality checks with Continuous Integration
    \item Jupyter/Rmd notebooks should be used for data analysis, but not code development
    \item Adopt a consistent development workflow and coding style for the project duration
    \item Publish research code with a DOI, citation file and software licence
    \item Optionally, explore synthetic data and executable papers to enhance reproducibility
\end{itemize}
\end{tcolorbox}

\section{Resources and Training} \label{training}

To make the most of the recommendations in this report, researchers will need to have gained some experience programming in a commonly used data science language such as Python or R, and have some familiarity with the basics of version control with Git. This section lists some useful online resources, reading material and links to organisations that run training courses. These resources will be useful to researchers who wish to learn the basics of research software engineering and data science skills, but also to more experienced researchers looking to expand their knowledge of these topics and practices.

\subsection{Turing Courses}

The Alan Turing Institute's \href{https://alan-turing-institute.github.io/rsd-engineeringcourse}{Research Software Engineering course with Python}:
\begin{itemize}
    \item Online resource enabling learning via Jupyter notebook tutorials
    \item Version Control section is language agnostic
    \item Topics include testing and advanced Python programming
\end{itemize}

The Alan Turing Institute's \href{https://alan-turing-institute.github.io/rds-course/}{Introduction to Research Data Science} is also in development (work in progress at the time of writing).

\subsection{Software Carpentry}

Visit \href{https://software-carpentry.org/}{software-carpentry.org} to find out about the wide range of lessons and workshops available.
\begin{itemize}
    \item Run by \href{https://carpentries.org/}{The Carpentries} in the United States
    \item Relevant \href{https://software-carpentry.org/lessons/}{lessons} include Version Control with Git and best practices for reproducible scientific analysis with R
\end{itemize}

\subsection{The Turing Way}

\href{https://the-turing-way.netlify.app/welcome}{Online handbook} developed by an open source community \citep{the_turing_way_community_2019_3233986}. Many of the topics in the \href{https://the-turing-way.netlify.app/reproducible-research/reproducible-research.html}{Guide for Reproducible Research} have been adapted to the TRE context for this report.

Additional topics covered by this resource not featured in this report include:
\begin{itemize}
    \item Code reviewing processes
    \item Research data management
    \item Risk assessment
    \item Project design
    \item Working collaboratively with GitHub
    \item Guidance for science communication
    \item Introduction to research ethics
\end{itemize}

\section{Acknowledgements} \label{acknowledgements}

With thanks to the WMML project researchers: Rowena Bailey, James Rafferty, Jane Lyons, Ashley Akbari (Swansea University), Farideh Jalali, Thamer Ba dhafari (University of Manchester), and PIs Ronan Lyons (Swansea University), Niels Peek (University of Manchester). With thanks also to May Yong, Emma Karoune and the Research Engineering Group at The Alan Turing Institute and The Turing Way community.

This work was supported by the Medical Research Council (MRC), grant no. MR/S027750/1.  This work was supported by Health Data Research UK, which receives its funding from HDR UK Ltd (HDR-9006) funded by the UK Medical Research Council, Engineering and Physical Sciences Research Council, Economic and Social Research Council, Department of Health and Social Care (England), Chief Scientist Office of the Scottish Government Health and Social Care Directorates, Health and Social Care Research and Development Division (Welsh Government), Public Health Agency (Northern Ireland), British Heart Foundation (BHF) and the Wellcome Trust. This work was supported by the ADR Wales programme of work. The ADR Wales programme of work is aligned to the priority themes as identified in the Welsh Government’s national strategy: Prosperity for All. ADR Wales brings together data science experts at Swansea University Medical School, staff from the Wales Institute of Social and Economic Research, Data and Methods (WISERD) at Cardiff University and specialist teams within the Welsh Government to develop new evidence which supports Prosperity for All by using the SAIL Databank at Swansea University, to link and analyse anonymised data. ADR Wales is part of the Economic and Social Research Council (part of UK Research and Innovation) funded ADR UK (grant ES/S007393/1).

\FloatBarrier
\def\UrlBreaks{\do\/\do-} 
\newpage
\bibliographystyle{apalike}
\bibliography{bibliography}
\footnotesize{Anaconda Software Distribution. Computer software. Vers. 2-2.4.0. Anaconda, Nov. 2016. Web. <https://anaconda.com>.}

\rfoot{
    \includegraphics[width=0.18\textwidth]{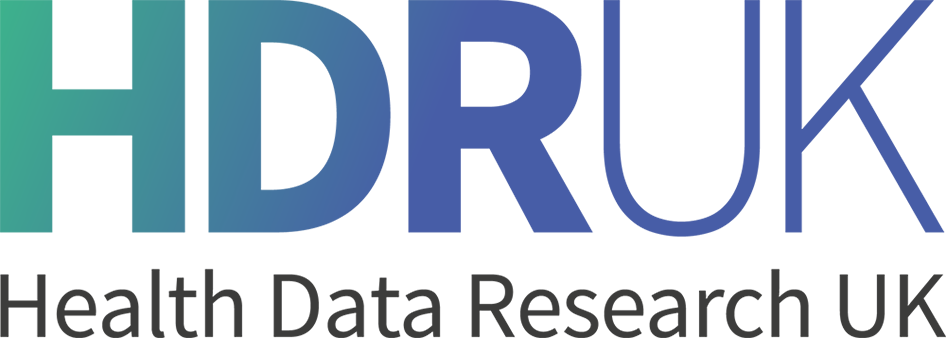}
    \includegraphics[width=0.23\textwidth]{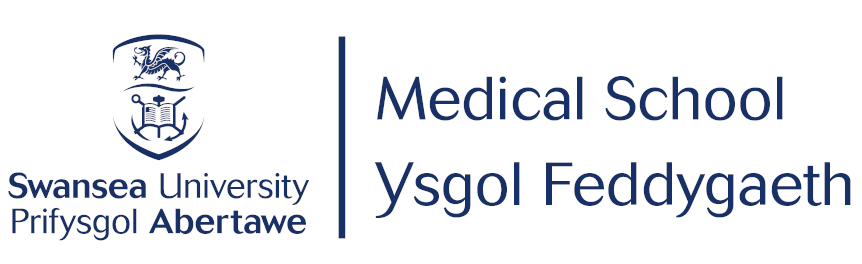}
    \includegraphics[width=0.18\textwidth]{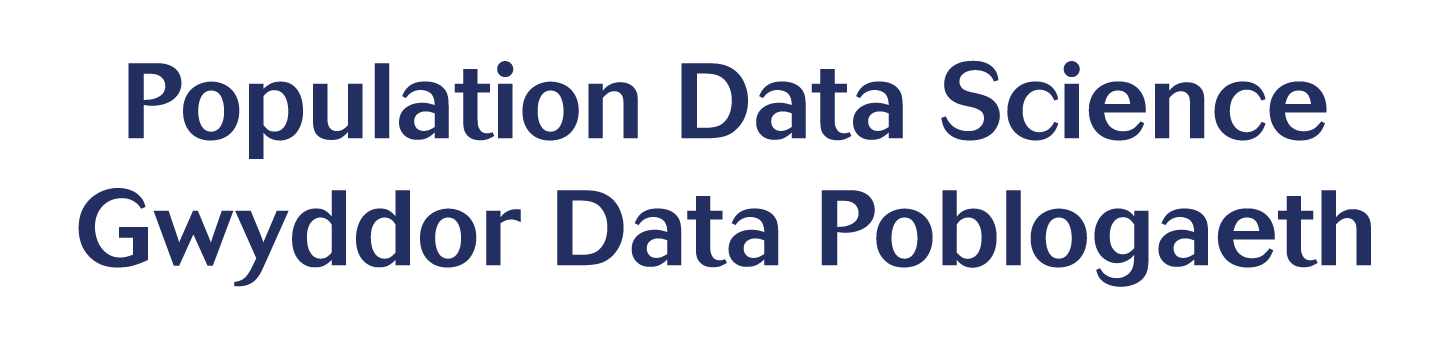}
    \includegraphics[width=0.18\textwidth]{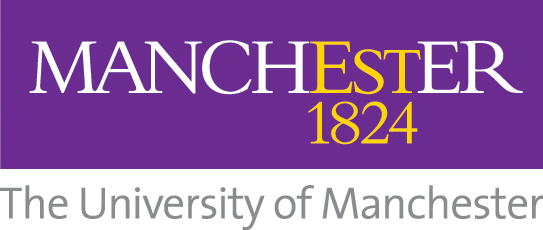}
    \includegraphics[width=0.18\textwidth]{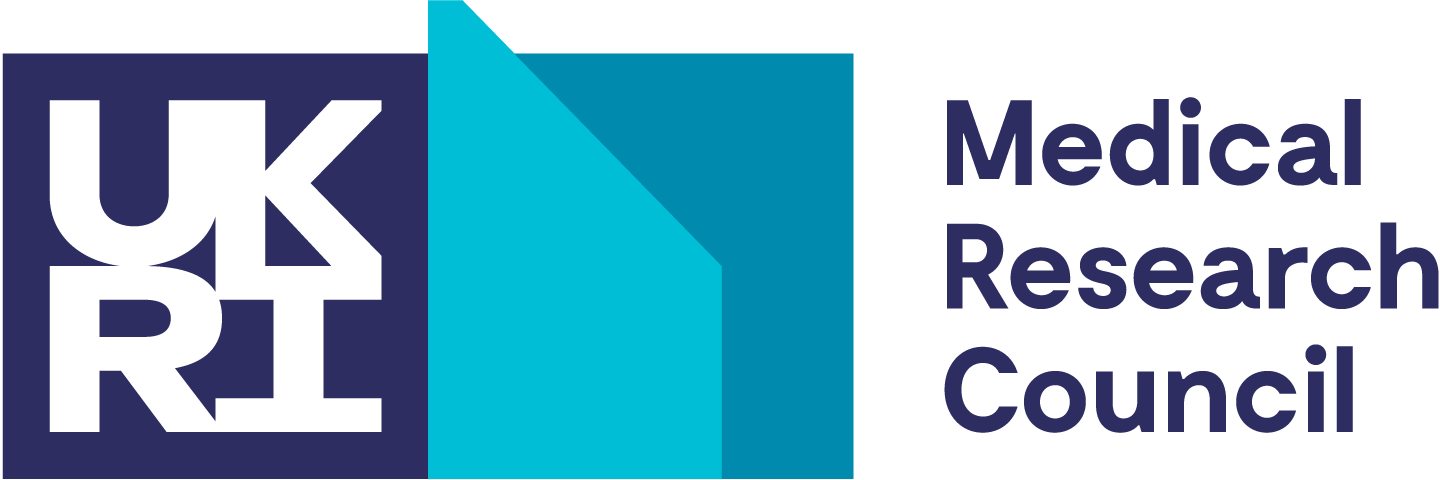}
    }

\end{document}